\documentclass[conference]{IEEEtran}
\IEEEoverridecommandlockouts
\usepackage[ngerman, english]{babel}
\usepackage{cite}
\usepackage{amsmath,amssymb,amsfonts}
\usepackage{algorithmic}
\usepackage{graphicx}
\usepackage{textcomp}
\usepackage{xcolor}
\def\BibTeX{{\rm B\kern-.05em{\sc i\kern-.025em b}\kern-.08em
    T\kern-.1667em\lower.7ex\hbox{E}\kern-.125emX}}

\newcounter{saveenumerate}
\makeatletter
\newcommand{\enumeratext}[1]{%
	\setcounter{saveenumerate}{\value{enum\romannumeral\the\@enumdepth}}
\end{enumerate}
#1
\begin{enumerate}
	\setcounter{enum\romannumeral\the\@enumdepth}{\value{saveenumerate}}%
}
\makeatother

\begin{document}

\title{Organic Computing as Chance for Interwoven Systems\\	
}

\author{\IEEEauthorblockN{Tobias Eckl}
\textit{University of Passau} \\
eckltob@fim.uni-passau.de}

\maketitle

\begin{abstract}
Systems are growing into more complex ones for developing and maintaining. Existing systems which do not have much in common on the first look are connected, due to the technical progress, even if it was never intended that way. 
It is an upcoming challenge to handle these large-scale and complex systems. A solution must be found to manage these ''Interwoven Systems''. Therefore it is discussed where approaches of ''Organic Computing'' can help, to handle some of these upcoming challenges.
\end{abstract}


\section{Motivation}
Information an Communication technology progresses very fast, the data produced and stored doubles about every year. \cite{b9}
We have a large quantity of data where we can get information. Data is collected everywhere and often communicated to various systems. This kind of systems already found their way into many parts of our life. 
 Once systems were designed and developed to do a single purpose and to work on their own, independently from other systems but more and more systems are connected with each other e.g., the ''smart'' systems which have spread into our homes.
Even so, these connected systems are still working on their own, to fulfill their main purpose, they also share information with each other and are connected in that way.
 If you take a look at these connections between the smaller systems you can also detect a larger system, consisting of systems, which is built of the connections of these smaller systems.
Even if all parts work independently from each other, they still have some influence on other parts of the system.
When time is progressing more and more systems will be connected with each other. The bigger the systems become more side effects can occur which were never intended. These side effects can cause lots of problems for the users, the system administrators, and for the developers designing parts for the system.
 The further you interweave systems with each other,  the bigger these problems can grow.  One question is when will these problems be too big or even too complex for us to handle.
  Before this happens something has to be done, so we do not lose control of these growing systems. The growth of complex systems due to interweaving and connecting more and more systems cannot be stopped, and the connections can also bring benefits.
 A solution is needed,  how to deal with these large-scale ''Interwoven Systems''. One approach is to let the system itself deal with upcoming problems, by increasing its self-organization. That way, the most upcoming problems should be managed by the system itself but without giving up the control over the system. 
    To increase the self-organization of the system one approach is, to use attempts of Organic Computing. This paper explains in what way the approach of  ''Organic Computing'' can control problems that result from such large-scale ''Interwoven Systems''.

\section{Interwoven Systems}
In the future, we will have to deal with large-scale systems. Those systems are growing even further, consisting of very different parts, so these systems are very heterogeneous.
 These components which are grown together still do not fully depend on each other. Some of the components are easily exchangeable.
  Such systems are growing over time, so there can be newly engineered parts together with older ones. Parts can be substituted, so the system is changing over time.\cite{b4}
In order to understand and define the term, ''Interwoven Systems'' (IS), the cornerstones this term is built on are introduced. That finally leads to a term-definition of the IS. 

\subsection{Cornerstones for Interwoven Systems}
As the basis for describing IS we start with the term ''System of Systems'' (SOS).
The term ''System of Systems'' is used describing a class of systems, but there is no commonly agreed definition of this term. The best known are the following:
 ''A system is a collection of entities and their interrelationships gathered
together to form a whole greater than the sum of the parts.''
\cite{b2}
and ''A system of systems is a set of different systems so
connected or related as to produce results unachievable by the
individual systems alone.'' \cite{b3}

\subsubsection{System of Systems}

Even so, these definitions are different there are major similarities.
The SOS is a large system consisting of heterogeneous parts forming this new more complex system.
 The SOS can be distinguished from other large systems, and the SOS can  be characterized by five fundamental characteristics \cite{b1}.
\begin{enumerate} 
	\item Operational Independence of the Individual System: 
    Each SOS consists of a set of smaller systems. These can act on their own, fulfilling their own goals, they do not depend on each other. They can operate independently. If you decompose the SOS into the component systems, each component still must maintain its own performance.
	
	\item Managerial Independence of the System:
	Within the SOS each component system is integrated and maintained independently. The existence of the SOS is not necessary for the component systems.
	
	\item Geographic Distribution:
	The component systems of SOS can be spatially distributed. The SOS is build of the component systems communicating with each other on various communication methods.
	
	\item Emergent Behaviour:
	Each component system works independently and has its own behavior. The behavior of the SOS resulting from the merger of the component systems is more than just an aggregation of the individual behaviors of the merged component systems. Novel functionality appears which cannot be achieved directly from the single components.
	
	\item Evolutionary Development:
	Every SOS is a not static system every component system can be replaced or removed. The structure, organization, and functionality of the SOS can change over time. Only limited external control is applied which leads to a self-evolutionary process.
	
\end{enumerate}

The SOS characterization covers a group of systems, but the challenges of upcoming systems will go far beyond this definition and characterization of the SOS.
 Communication mechanisms allowing the component systems to interact with each other have to be established. The systems must learn to exchange understandable problem descriptions for collaboration in the sense of application-oriented information and automatically deal with such problems.
This might be realized by dynamic ontologies since component systems, and challenges change at runtime.

\subsection{Federations of Systems}
To fulfill common objectives or complex missions often large sets of systems are needed to reach these goals. With large sets, it is often impossible or at least very limited to achieve a synchronized and coordinated behavior of the involved SOS, because centralized control of the component systems is not possible or very limited \cite{b4}.
In such cases  a ''Federation of Systems'' is needed (FOS – see \cite{b1} )
to achieve the goals better.
The term FOS has been defined by Krygiel as an SOS ''[...] managed
without central authority and direction.'' \cite{b3}
This means the component systems of a FOS are completely independent, working on their own pursuing their own goals. Without the centralized control over the SOS, all parts have to collaborate and cooperate that the FOS within the self-organization process can come to a decision \cite{b4}. 
A FOS therefor can be characterized by three key aspects:
\begin{enumerate}
	\item a high degree of autonomy
	\item heterogeneity in terms of the participating SOS
	\item distribution of organization structure and processes
\end{enumerate}

Overall the FOS cannot cover all kind of upcoming systems. Without a hierarchical structure in a FOS there exist many goals which can be inconsistent and be conflicting such complex tasks require a problem specific structure and organization of the system.
 Such structures cannot be static because the participating SOS can chance very easy, so the hierarchies have to be established by the systems themselves. They also may be terminated or adapted due to the changing system in conditions, requirements, and goals.

\subsection{Interwoven Systems}
The terms SOS and FOS are used to define a concept to describe current technical systems, but not all systems are covered.
The characterizations are building a basis for describing even more complex systems.
\subsubsection*{Characterization of Interwoven Systems}
The Term Interwoven Systems (IS) builds upon the previous characterizations and definitions of SOS and FOS. 
So Interwoven Systems consist of a set of component systems which can be independent.
Each component system can also have its own component systems.
Therefore, the previous five characteristics of SOS remain valid for the IS but are further augmented by a set of important characteristics \cite{b4}.

\begin{enumerate}
	\item Operational Independence of the Individual System:
	There cannot be a centralized control for an IS, because the IS has changing component systems and has various administrative authorities, which can have different goals. Interwoven Systems are characterized by self-organization of systems and their federation.
	
\item  Managerial Independence of the System: 
Component systems can belong to the same authority, but they still have to be handled as individual systems, that the IS can easier be maintained. IS can be characterized by changing administrative domains.

	\item  Geographic Distribution:
 Every IS consists of component systems which are building a set of interconnected systems. The component systems can be distributed, but a geographical distribution is not strictly necessary, large data centers can also be understood as IS then the geographical distribution is limited to a smaller area.
 From a data perspective, synchronization of data reflecting the time-variance of IS is needed. 
	 Such an IS is characterized by the possibility to separate the component systems in the sense of defining system boundaries as administrative domains on the basis of geographical separation.
	 
	\item  Emergent Behavior:
	Emergent effects can either be positive or negative. The emergent behavior can result from the self-organization and the interaction between component systems. 
    By coupling independent systems and building a larger IS new tasks can be handled which also can lead to unanticipated new behavior.
IS should recognize such emergent behavior and act accordingly without external intervention.

	\item  Evolutionary Development:
	Typically IS are changing over time at runtime they can grow but component systems also can be removed. So it is not possible to manage the system by one authority. 
	 The system is allowed to adapt itself to changing conditions by self-organization, but then design-time decisions have to be made at runtime and in the responsibility of the individual system. An IS is characterized by a continuous development where no user interaction is needed, and the IS manages its development by itself.
	
\enumeratext{Additionally to the previous characteristics which are based on the SOS characterization, the term IS goes further by also in networked nature and distributed self-organized management and control of this kind of systems. The following characteristics go far beyond the SOS description.}

\item  Mutual Influences of Networked Systems:
	Individual component systems can organize and optimize itself due to the self-organization of the component system as part of an IS. The adaption of a component system can have mutual influences:
The component systems are all coupled, the adaption of the component system can cause the need for other component systems to adapt themselves to the changes. This adaption easily can lead to an uncontrollable and oscillating behavior of the systems.
An IS must prevent all kinds of uncontrollable behavior by itself. There is a need for a federative approach and a kind of smartness within the cooperation and coupling between the component systems.

\item  Heterogeneity of Component Systems and Federations:
	The contained component systems of an IS can be very different so there is is a large heterogeneity, which comes from the openness of the networked system. 
	Component systems are individually designed and are working on their own goals, so they can influence each other quite easily.
	 Since there is no central authority for managing the component systems, the component systems can have an unpredictable influence on other partners of the collaboration.
	  In the worst case, they even can have a malicious influence on other parts of the IS or the entire system even this was never intended by the developer of the component system. 
	An IS is characterized by security and trustworthy mechanisms which allow the IS to deal with these problems without giving up the openness and heterogeneity.
\item  Uncertainty:
Due to the heterogeneity the self-organization, the self-organized adaption of component systems and the continuous evolution of such systems the state and the behavior of the IS is not fully predictable.
So an IS is characterized by uncertainty in the system's behavior.

\end{enumerate}

\subsubsection*{Term Definition: ''Interwoven Systems''}
When the previously defined eight characteristics are fulfilled, a system can be referred to as an Interwoven System \cite{b4}. An IS is consisting of component systems which are coupled and interacting with each other, they also communicate directly or indirectly.
 An IS is an ultra large scale system which can handle changes of itself at runtime which were not defined or intended at design time. The individual component systems can change in architecture, parameter, and goals.
At the same time the overall IS can also change, e.g., component systems can be added or removed, communication, infrastructure, the logical structure of the collaboration between component systems, and even the set of goals to be achieved by the IS.
   The component systems are independent and working on their own, but the behavior of the is is not only an aggregation of the functions of the component systems.
Additionally, the conflict between different goals has to be avoided there are a large functional repertoire and the interaction between the component systems.

Challenges for IS are optimizations on how the component systems can collaborate and how the situation the system is can be modeled \cite{b5}.
 Further, the IS can never be in an unstable state so the component systems must have a balanced and goal-oriented behavior for the IS. The management of the system needs to handle this class of systems and to maintain the desired behavior and performance.

\section{Examples of Interwoven Systems}
\subsection{Power Management Systems}
Today’s power management systems (PMS) fulfill already the characterization of a SOS, but future PMS will change towards IS with the previous characterizations (1 - 8) \cite{b4}. The PMS considered as a whole system consists of a large number of power plants.
 They are connected, but each power plant still can operate on its own, independently from the others (1 Operational Independence).
  Each of the power plants is managerial independent (2), to a certain point the power plants are working economically independent from each other.
  The power plants are naturally geographical distributed (3), the distribution is even increasing, due to the process of building smaller power plants such as solar plants, wind farms.
  The stability of the PMS is an emergent behavior (4) since no power plant can provide stability or amount of power needed for the PMS on its own. So the PMS fulfills a task, a single part of the system cannot handle.
   Old power plants can be shut down, and new power plants are taking their places, the technology is also changing, so there is evolutionary development (5) in the PMS.
Due to the growing number of generators, the increasing
dependence on unreliable sources (wind, solar), and the increasing
ability to control distributed energy resources, the
properties (6) to (8) of IS will become a necessity to master in
future PMS. \cite{b4}
Other SOS such as gas distribution grids, or district heating systems can be connected to the PMS, so an even larger system can be built. 
The combination and coupling of these resulting SOS
remain very complex and unpredictable. The coupling of
several SOS and networks lead to new challenges. 
 With increasing numbers of unreliable sources for the PMS, e.g., wind and solar energy there might be higher requests for energy on other parts of the PMS. If we send the request to a combined heat and power plant, this may cause a conflict.

\subsection{Vehicular Traffic}
Traffic control and management is an example of an already existing IS \cite{b4}. 
To control traffic lights, there are strategies depending on the observed traffic conditions, in term of vehicles passing the underlying intersection. The intersection controller has always the task to optimize the traffic flow. This task varies in terms of the controlled intersection's topology, and the controller's position in the network. 
 Residential areas, arterial roads, and highways have different requirements for the controller. Some controllers cannot only set up the green times of the traffic lights but also determine speed limits, in cooperation with other controllers they can make up strategies to optimize the traffic flow of a larger area and relieve intersections by guiding drivers through the network recommending routes.
If we also take other carries, like aircraft, railways, and pedestrians into account the traffic system becomes even more heterogeneous. The intersections and the carries belong to complete different authorities, but the resulting system has a broader scope than just looking at one of the parts.
 There are goals for this system which is looking at the whole traffic system such as minimizing pollutions, waiting times for travelers. 
Trying to reach these goals even in a small closed environment is difficult, cause there are always effects from neighboring network parts.
 We have an IS in the traffic control and management systems consisting of heterogeneous entities in different authorities.

\section{Challenges}
In order to handle such large-scale systems, there are various challenges.
Systems that go into production are only slightly adapted and updated \cite{b5}. So increasing the degree of self-organization means not only to allow to adapt self-properties, like self-configuration, self-optimization or self-healing.
 It also demands concepts for changing and managing the structure and composition of large-scale integrated systems at runtime and without the help of a user.
  So developed component systems need the abilities to fulfill their role in place within a large scale integrated system at runtime independent without the help of a user.
The systems must manage their place within the large-scale coupled systems, and even when needed autonomous modify their design. They also need to reflect on the behavior of connected systems just like self-reflecting own behavior.
The large IS do not have a single global function. Many criteria have to be fulfilled in the system, so optimization is very complex. The optimal behavior for single functions will not lead to a good result for the global system, so the component systems need to manage these problems and find their way for a global result.
To get good results for the optimization, these are some key characteristics \cite{b5}:
\begin{enumerate}
	\item Optimization is context-dependent. To decide if a solution or a step towards a solution is good depends on the context. Mostly it is a compromise, so it includes dropping goals or even giving up on an objective, or not to persist on a single goal. So the optimization cannot be fixed on simple and unalterable criteria.
	
	\item As noted above, the criteria for optimization cannot be fixed.
	The system always can change, e.g in the component systems or the goal of the whole system may vary a bit. Hence, the optimization for the system is continually changing.
	
	\item There are always more ways to solve a problem, and the best way differs on the combination of objectives and how resources are adjusted. There is no best way because the requirements are varying.
	
	\item  The system(s) must always be continually viable, that means it is in a good enough state to survive, satisficing, in a good safehold if an emergency state occurs, etc..Interwoven Systems cannot go totally offline. Not making decisions at all can cause problems, and can have severe consequences for the whole system.
	
	\item Goals can be modified at runtime this results in changes for the system the component systems have to adapt.
	 Such a change can be triggered by users but also by the system itself. Consider for instance in the traffic control: the normal goal is to decrease averaged waiting times.
	 In the case of oversaturated situations, the system might switch to the goal of relieving as much traffic as fast as possible.
	These are contradictory strategies. We need concepts to
	allow for such a change and for techniques that can adapt
	quickly and reliably.
	
	\item Optimization at this system level can include tradeoffs
	among goals, and this includes issues close to
	scheduling problems and social issues such as trust.
	Component systems must work on suboptimal levels for their own goal to support the global one as long as they achieve theirs. At this
	level, there can be new strategies and new policies.
\end{enumerate}
As one can see, evaluation, understanding, and defining
what is good or satisfying or optimal in Interwoven Systems,
is complex.
 Instead of finding the optimal solution, we need
good enough and fast enough results. So the system is always in a stable state.
These are major challenges which cannot be mastered easily with traditional methods which are reaching their limit, further is discussed how the approach of ''Organic Computing'' will be able to deal with these challenges.

\section{Organic Computing}
\subsection{Characterization}

The adjective ''organic'' has several meanings in The Cambridge Dictionary. The meaning which is relevant for the term ''Organic computing'' is: ''It aims at augmenting technical systems with properties that are similar to those found in living things.''
With Organic Computing (OC) artificial systems should get a behavior which is more oriented on living things \cite{b8}. 
To get such behavior, the systems must be aware of their surroundings and their environment.  The number of large autonomous systems which are equipped with sensors and actuators increases.
In order to get a behavior which is oriented on living things, the systems need to communicate freely and organize themselves to perform actions and services. So in \cite{b7} OC ''...is a technical system, which is equipped with sensors to perceive its environment and actuators to manipulate it. It adapts autonomously and dynamically to the current conditions of the perceived environment. This adaptation process has an impact on the system's utility, which is continuously improved by the organic system itself.'' 

In order to react to previously unknown and unanticipated conditions, with appropriate behavior, an organic system is typically based on (machine) learning techniques.

To reach this kind of behavior, the so-called self-* mechanisms are required \cite{b7}.
\begin{enumerate}
	\item Self-configuration is used to modify the parameters of the system.  Organic systems can configure themselves by adjusting the parametrization, resulting in different behavior, in order to adjust their behavior to higher-level user goals.
	
	\item Self-organization helps the system to adapt to changes in the structure of component systems or the connection. This adaption has to be done continuously at runtime based on the active user goal, and the status of the changes has to be communicated to the other subsystems.
	
	\item Self-integration is related to self-configuration and self-organization in the context of combination and interactions of several systems.
	The organic system decides autonomously about its role within the whole system and adapts its behavior as well as the relations to other parts.
	
	\item Self-management includes self-configuration, self-organization, and perhaps further self-* mechanisms.
	
	\item Self-healing is the ability of the organic system to detect, diagnose, repair failures and even localize the failure.
	
	\item Self-protecting to protect the organic system as a component system, as well as the whole system from attacks from outside. This protection goes also on large-scale failures which cannot be handled by the self-healing aspect.
	
	\item Self-stabilizing guarantees that an organic system is always in a stable state which results in a stable behavior of the system, even when the state of the system is changing over time. 
	
	\item Self-improving means that the organic system continuously analyses its decisions so even when there was already a solution with the altering system it is possible, that better solutions can be found. A machine learning mechanism is needed to fulfill this goal, to improve the system at runtime.
	
	\item Self-explaining so the user always can keep the control over the system. The system needs a high degree of autonomy to fulfill its purpose.
	
\end{enumerate}
An organic system can consist of a large set of autonomous subsystems. But each OC system has two complementary parts \cite{b7}:
\begin{itemize}
	\item  the first is responsible for the productive operation of the system	(i.e, it fulfills the technical purpose)
	\item the second is for the adaptation aspects (i.e., realizing the organic capabilities).
\end{itemize}

\subsection{Organic Computing as chance}
As described in the previous sections, IS are giving us challenges we need to solve, and it seems that OC is giving us the possibility to master some of these problems.
 It seems possible to design and build an IS as an organic system. IS produce a large amount of data, the component systems are getting information out of the data but since there is no centralized control some information of the data other component systems want might not be shared.
 In an organic system with self-configuration parameters can be changed and additional information can get to the whole system. 
Since in an IS the structure of component systems can change at runtime these changes of the IS can even be handled easier with the self-organization and self-integration aspect of an organic computing system.
Due to this changing structure of the IS, there can occur failures which can be handled by the self-healing of an organic computing system.
The self-protection goes against malicious attacks as well as against bigger failures that can occur in every system.
An IS always has to be in a stable state to produce results, with the self-stabilizing aspect such a state can always be guaranteed.
The optimization in an IS has various challenges, as described in (IV), these challenges can be mastered if the system is self-improving, so even when the system has not the optimal solution, the results can get better over time. This is a good approach for the optimization issues we face in an IS.

\section{Conclusion}

The term IS is describing the many challenges we need to face.  OC, on the other hand, is a concept giving answers to these challenges we face from the IS. There is still a big step from the theoretical approach of solving the challenges to actually designing and creating such organic computing systems which can handle Interwoven Systems.  Designing and creating new systems considering these concepts from the start should be the first and easier step. The so created systems can grow when all component systems are designed to work together. 
Most of the Interwoven Systems are not created new but growing together from single independent systems with technology from different ages. 
Looking at the PMS power plants are substituted, but it is a long and expensive task which cannot be done overnight. There are also many companies with different interests, which all have to agree on the organic administration. Organic Computing is not the answer for all IS, but surely will help to handle some.
The challenge to adjust these systems needs much work.
Managing this kind of IS will be a great challenge for the future.

\end{document}